\documentstyle[twoside,fleqn,espcrc2,epsf]{article}

\title{
Scaling in SU(3) theory with a MCRG improved lattice action
\thanks{
Poster presented at Lattice '97 by A. Bori\c{c}i}
}

\author{
A. Bori\c{c}i and R. Rosenfelder
\address{
Paul Scherrer Institute, CH-5232 Villigen PSI, Switzerland}
}
       
\begin{document}

\begin{abstract}
We test various improved gauge actions which are made of
linear combinations of Wilson loops. We observe the restoration
of rotational symmetry in the static interquark potential already
on coarse lattices as small as $6^312$. Furthermore, we study scaling and
asymptotic scaling of the string tension with a MCRG-improved
action on $12^324$ lattices. Preliminary results show that scaling
sets in at $a \approx 0.3 $ fm.
\end{abstract}

\maketitle

\section{INTRODUCTION}

There is a great deal of work being done to reduce artifacts
in lattice QCD \cite{review}.
This is important, if one aims to extract precise continuum
physics on the emerging teraflop-range computers.

We will focus here on the improvement of the pure gauge sector
\cite{LW,IW,Alford_etal,TARO}. The general form of the action that we try is
\begin{equation}
\begin{array}{l}
S_{\rm gauge} = - \beta \sum_{\rm loops} \Bigl ( \> \mbox{Square} \\
          + c_1 \cdot \mbox{Rectangle} + c_2 \cdot \mbox{Parallelogram} \> 
\Bigr)
\end{array}
\end{equation}
For $c_1 = c_2 = 0$ one recovers the Wilson action. The additional Wilson
loops are six-link loops in two and three dimensions (excluding the
bent rectangle which is redundant to order $O(a^2)$).

A ``good'' improved action should reduce lattice artifacts and should 
be practical for computations. Therefore, one would like to have as little as 
possible additional terms in the action.
Classically, at order $O(a^2)$, a convenient choice is
$c_1 = - 1/20$ and $c_2 = 0$ \cite{LW}. Corrections of these values at
leading order in bare coupling amount to a small nonvanishing $c_2$.
Recently, it was observed that an expansion in a renormalized coupling
\cite{LM} produces an impressive improvement of the classical values.

A different strategy to select couplings is to compute them by performing
Renormalization Group (RG) transformations. Ideally, one could construct
Fixed Point (FP) or ``perfect'' actions \cite{PHasenfr}. For computational
convenience, we prefer RG improved actions which are less complex
\cite{Pennanen}.
One earlier approach has
blocked continuum gauge fields by expanding in the bare coupling
\cite{IW}. More recently, by blocking lattice gauge fields during a
Monte Carlo simulation (MCRG), one has constructed improved gauge actions
\cite{TARO,Takaishi}.

Since a thorough study of scaling with the MCRG improved gauge actions is
missing, we decided to investigate this problem.
(The corresponding study with the Iwasaki action
can be found in \cite{Itoh_etal,Iwasaki_etal}.)
As we will see RG improved actions lead to a better
restoration of rotational symmetry 
than mean field improved actions. Therefore we decided 
to study also the scaling properties with the MCRG improved action.

\section{ROTATIONAL SYMMETRY ON COARSE LATTICES}

In the table below we give the simulation parameters with improved
gauge actions on small lattices.

\vspace{-1cm}
\begin{table}[htb]
\begin{center}
\begin{tabular}{|l|c|l|c|l|}
\hline\hline
Action & $c_1$ & $~~~c_2~~~$ & $\beta$ & Lattice size\\
\hline
MF1       &  -0.074~  &  0        &  6.25 & $~~~6^312$ \\
MF2       &  -0.0827  &  -0.0124  &  6.80 & $~~~6^4$   \\
IW        &  -0.0907  &  0        &  6.50 & $~~~6^312$ \\
MCRG      &  -0.115~  &  0        &  7.20 & $~~~6^312$ \\
\hline\hline
\end{tabular}
\end{center}
\caption{Simulation parameters on coarse lattices}
\end{table}
\vspace{-.5cm}

The couplings of the mean field improved action are given by tadpole improved
one loop results (for their definition see \cite{Alford_etal}).
For all the actions we generated 100 independent configurations,
out of 11000 sweeps in total. More results with increased 
statistics will be reported  elsewhere \cite{Bor_Ros}.

In Figure 1 we show the static potential in lattice units for the
above actions. The solid line
is a 3-parameter
fit: $V_0 + K r - \alpha/r$ with the
scale set by the string tension
$\sqrt{\sigma} = 420$ MeV.
\begin{figure}
\centerline{\epsfysize = 2.5 in \epsffile {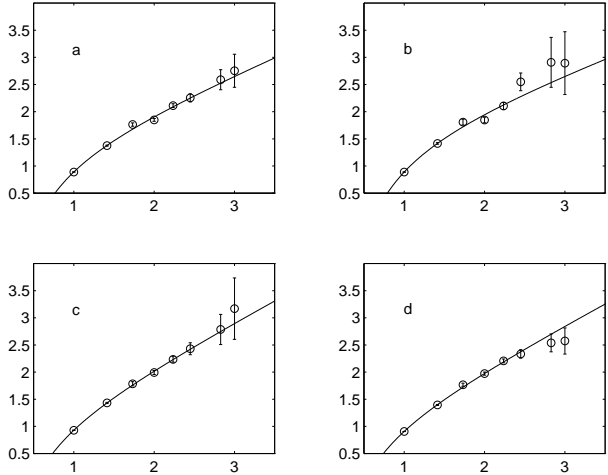}}
\vspace{-1cm}
\caption{Static interquark potential on coarse lattices
with MF1 (a), MF2 (b), Iwasaki (c) and MCRG improved actions (d)}
\end{figure}
\begin{figure}
\vspace{-1.5cm}
\centerline{\epsfysize = 2.8 in \epsffile {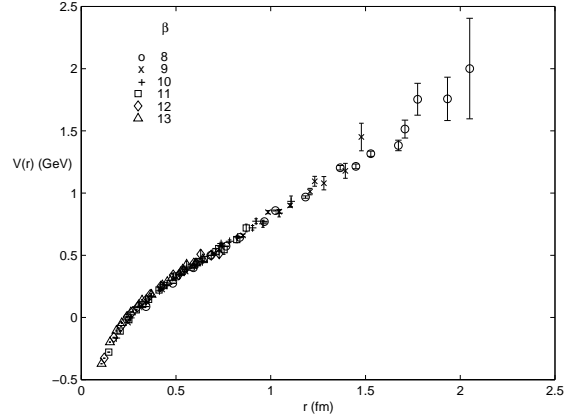}}
\vspace{-1cm}
\caption{Scaling of the static interquark potential on $12^324$ lattices
with the MCRG improved action}
\end{figure}
In  Table 2 we give the quality of the 
fit and the lattice spacings.

\vspace{-1cm}
\begin{table}[htb]
\begin{center}
\begin{tabular}{|l|c|c|}
\hline\hline
Action & $ \chi^2/dof $ & Lattice spacing (fm) \\
\hline
MF1       &  0.24  &  0.37(3)  \\
MF2       &  0.67  &  0.35(6)  \\
IW        &  0.07  &  0.42(1)  \\
MCRG      &  0.07  &  0.42(2)  \\
\hline\hline
\end{tabular}
\end{center}
\caption{Quality of the fit to the potential parametrization and  the
corresponding lattice spacings on coarse lattices}
\end{table}
\vspace{-.8cm}
Our results indicate that the Wilson action should not be used in simulations
on coarse lattices. We observe further that violations of rotational
symmetry are smaller for the RG improved actions.

\section{SCALING OF STRING TENSION WITH THE MCRG IMPROVED ACTION}

We have performed simulations on $12^324$ lattices with the MCRG
improved action for
$\beta \in [8,13]$ with increment $\Delta\beta = 0.5$.
The main results are tabulated below (see \cite{Bor_Ros} for details).

\vspace{-1cm}
\begin{table}[htb]
\begin{center}
\begin{tabular}{|r|c|c|c|c|}
\hline\hline
$\beta~~$ & $K$ & $\chi^2$/dof & $a$ (fm)\\
\hline
 8.0  &  0.5147(103)   &  0.4901  &  0.3416(34) \\
 8.5  &  0.3772(111)   &  0.1356  &  0.2924(43) \\
 9.0  &  0.2677(59)~   &  0.083~  &  0.2464(27) \\
 9.5  &  0.1995(50)~   &  0.0521  &  0.2127(27) \\
10.0  &  0.1502(40)~   &  0.0056  &  0.1824(24) \\
10.5  &  0.1138(25)~   &  0.0022  &  0.1606(18) \\
11.0  &  0.0928(23)~   &  0.0011  &  0.1451(18) \\
11.5  &  0.0761(13)~   &  0.0002  &  0.1314(11) \\
12.0  &  0.0649(24)~   &  0.0007  &  0.1213(23) \\
12.5  &  0.0554(20)~   &  0.0004  &  0.1121(20) \\
13.0  &  0.0505(9)~~    &  0.0001  &  0.1070(10) \\
\hline\hline
\end{tabular}
\end{center}
\caption{Simulation results on a $12^324$ lattice and quality of
the fit for the static potential}
\end{table}
\vspace{-.5cm}

In Fig. 2 we observe scaling of the potential already at $\beta = 8$, our
starting point, corresponding to a lattice spacing $a \approx 0.34 $ fm.
 
Provided the fermion action is improved at the same order,
this result shows that one could compute the spectrum in quenched QCD
already on a $8^3$ space volume amounting to a lattice
size $L \approx 2.7 $ fm.
This conclusion is
consistent with the results of hadron spectroscopy on $8^316$
blocked configurations
used to determine the MCRG improved action \cite{Bor_For}.
Note that the same accuracy with the Wilson gauge and fermion action has been
achieved on $32^4$ lattices.

{\it
Therefore, these results together
with those in \cite{Bor_For} strongly suggest that one could perform
phenomenological lattice calculation
runs in the quenched approximation with sufficient
accuracy ($1\%$ level) on lattices with $12-20$ points in each dimension.
}
\begin{figure}
\vspace{-.3cm}
\centerline{\epsfysize = 2.2 in \epsffile {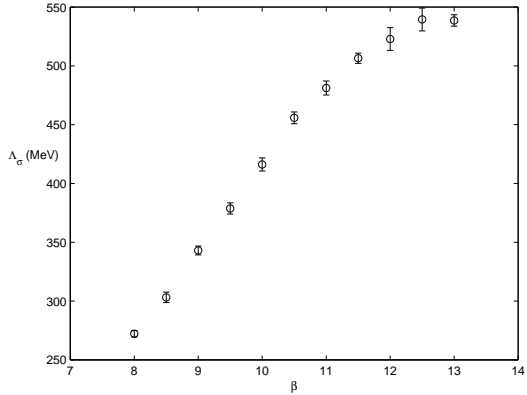}}
\vspace{-1cm}
\caption{$\Lambda_{LQCD}$ as measured by the string tension}
\vspace{-0.7cm}
\end{figure}

\begin{figure}
\vspace{-.3cm}
\centerline{\epsfysize = 2.2 in \epsffile {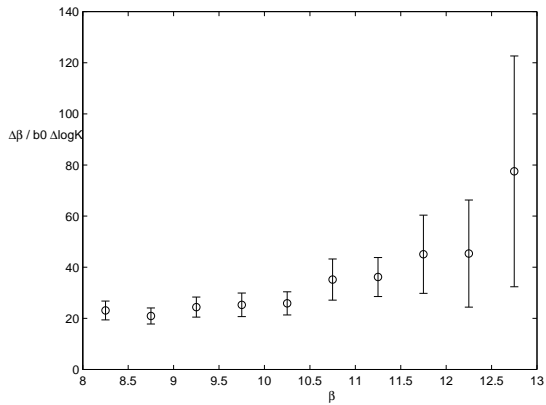}}
\vspace{-1cm}
\caption{Discrete $\beta$-function as a function of $\beta$}
\end{figure}
We have tested asymptotic scaling too. In Figure 3 we show
$\Lambda_{LQCD}$ as a function of $\beta$. It can be seen that a saturation 
seems to start already at $\beta = 12.5$.
In Figure 4 we plot the discrete beta-function based on
the string tension results. The results obtained up to now are 
not yet conclusive for the onset of asymptotic scaling.
At present we are running simulations at higher $\beta$s
on $18^336$ lattices
and at the same time we are increasing the statistics. The results will
reveal the behavior of the discrete $\beta$-function deep in the
perturbative region \cite{Bor_Ros}.

\vspace*{0.6cm}
We would like to thank F. Niedermayer for his helpful
comments.

\end{document}